\documentstyle[11pt,aaspp4,tighten,flushrt]{article}

\newcommand{\kms}{\mbox{${\rm\,km\,s}^{-1}$}}
\newcommand{\pc}{\mbox{$\rm\,pc$}}
\newcommand{\kpc}{\mbox{$\rm\,kpc$}}
\newcommand{\msun}{\mbox{$\,M_\odot$}}
\newcommand{\microm}{\mbox{$\rm\,\mu m$}}
\newcommand{\mm}{\mbox{$\rm\,mm$}}
\newcommand{\ang}{\mbox{$\rm\,\AA$}}

\newcommand{\ie}{{\it i.e.,\/}}
\newcommand{\beq}{\begin{equation}}
\newcommand{\eeq}{\end{equation}}

\newcommand{\secspt}{$\buildrel{\prime\prime}\over .$}
\newcommand{\hst}{{\it HST\/}}

\lefthead{Statler, King, Crane, \& Jedrzejewski}
\righthead{Double Nucleus of M31}

\begin{document}

\title{Stellar Kinematics of the Double Nucleus of M31\footnotemark}

\footnotetext{Based on observations with the NASA/ESA {\em Hubble Space
Telescope\/}, obtained at the Space Telescope Science Institute, which is
operated by AURA, Inc., under NASA contract No. NAS 5-26555.}

\author{Thomas S. Statler}
\affil{Department of Physics and Astronomy, 251B Clippinger Research
Laboratories, Ohio University, Athens, OH 45701, USA; tss@coma.phy.ohiou.edu}
\author{Ivan R. King}
\affil{Astronomy Department, 601 Campbell Hall, University of California,
Berkeley, CA 94720-3411, USA; king@glob.berkeley.edu}
\author{Philippe Crane}
\affil{European Southern Observatory, Karl-Schwarzschild-Strasse 2,
 D-85748 Garching, Germany;  and Department of Physics and Astronomy,
6127 Wilder Laboratory, Dartmouth College, Hanover, NH 03755-3528, USA;
phcrane@vermontel.net} \and
\author{Robert I. Jedrzejewski}
\affil{Space Telescope Science Institute, 3700 San Martin Drive,
Baltimore, MD 21218, USA; rij@stsci.edu}

\vskip -5.3in {\hfill \sl To Appear in The Astronomical Journal, 1999 February}
\vskip 5.2in

\begin{abstract}

We report observations of the double nucleus of M31 with the $f/48$ long-slit
spectrograph of the \hst\ Faint Object Camera. We obtain a total exposure
of 19,000 sec.\ over 7 orbits, with the $0\farcs063$-wide slit along the line
between the two brightness peaks (PA $42 \arcdeg$).  Careful correction
of the raw data for detector backgrounds and geometrical distortion is
essential. A spectrum of Jupiter obtained with the same instrument is used
as a spectral template to extract rotation and velocity dispersion profiles
by cross-correlation. The rotation curve is resolved, and reaches a
maximum amplitude $\sim 250 \kms$ roughly $0\farcs 3$ either side of a
rotation center lying between P1 and P2, $0\farcs 16 \pm 0\farcs 05$ from
the optically fainter P2. We find the velocity dispersion to be $\lesssim
250\kms$ everywhere except for a narrow ``dispersion spike'', centered
$0\farcs 06 \pm 0\farcs 03$ on the anti-P1 side of P2, in which $\sigma$
peaks at $440 \pm 70 \kms$. At much lower confidence, we see local
disturbances to the rotation curve at P1 and P2, and an elevation in
$\sigma$ at P1. At very low significance we detect a weak asymmetry in the
line-of-sight velocity distribution opposite to the sense usually encountered.
Convolving our $V$ and $\sigma$ profiles to CFHT resolution, we find good
agreement with the results of Kormendy \& Bender \markcite{KoB99}(1999),
though there is a 20\% discrepancy in the dispersion that
cannot be attributed to the dispersion spike. Our results are not
consistent with the location of the maximum
dispersion as found by Bacon et al.\ \markcite{Bac94}(1994). Comparing
with published models, we find that the sinking star cluster model of
Emsellem \& Combes \markcite{EmC97}(1997) does not reproduce either the
rotation curve or the dispersion profile. The eccentric disk model of
Tremaine \markcite{Tre95}(1995) fares better, and can be improved
somewhat by adjusting the original parameters. However, detailed
modeling will require dynamical models of significantly greater realism.

\end{abstract}


\section{Introduction\label{s.introduction}}

The center of our neighboring spiral galaxy in Andromeda (M31 = NGC 224) 
has intrigued astronomers for many decades.  Although the usual 
overexposed portrait images of ``the Great Spiral'' 
give an impression of great blandness across the entire central region, 
this is far from the truth.  Visual observers at least as far back as
Lord Rosse recognized the abrupt rise in surface brightness
toward the center, to a peak that is unresolved even in the best 
ground-based seeing (Hodge \markcite{Hod92}1992 and references therein).
The photometric structure became still more intriguing when
Stratoscope II showed the nucleus to be asymmetric, with its
brightest point off-center with respect to the rest of the inner bulge
(Light, Danielson, \& Schwarzschild \markcite{LDS74}1974).

Stratoscope
resolution remained unsurpassed until two decades later, when Hubble Space
Telescope (\hst) imaging showed that M31's central brightness peak was actually
double (Lauer et al.\ \markcite{Lau93}1993).  In addition to a sharp
peak which is clearly at the center of
the isophotes farther than 1 arcsec from the center, there
is a second peak with a resolved core, about half an arcsecond
from the central peak along the major axis of this central
region.\footnote{The spatial scale is $1\arcsec = 3.5\pc$ for a distance
of $725\kpc$ (Hodge 1992).}
Because the eccentric 
peak was the brighter one in the $V$ band in which Lauer et al.\
observed, they named it P1, and the more central peak P2. This turns out
to be a somewhat unfortunate choice, for reasons we discuss below.

Efforts to map the rotation, velocity dispersion, and higher moments of
the line-of-sight velocity distribution (LOSVD) near the center of M31
have led to further puzzles. Given the known brightness peak, perhaps it
should have come as less of a surprise when Lallemand, Duchesne, \& Walker
\markcite{LDW60}(1960) discovered a rapid central rotation, which peaked
at $\pm 90 \kms$ at $3\arcsec$ on either side of the center.  Later
observations, profiting from better seeing and greatly improved
instrumentation, have sharpened the rotation curve, to $160 \kms$ at
$r\approx 0\farcs7$, and measured a central dispersion that approaches
$300 \kms$ in the best seeing (Kormendy \& Richstone \markcite{KoR95}1995).
Dynamical modeling of these data now points to
the M31 nucleus as one of the more likely sites for a supermassive black hole
(Kormendy \& Richstone \markcite{KoR95}1995 and references therein).

Recently, Bacon et al.\ \markcite{Bac94}(1994) used the TIGER multi-object
spectrograph, which records 400 spectra at closely spaced points, to
map the kinematic fields in two dimensions. They
reported that the velocity dispersion peaks at a point that coincides 
with neither brightness peak. The highest velocity dispersion that they found
was at a point 0\secspt7 from P2, in the direction away from P1.
This surprising result disagrees, however, with long-slit spectroscopy
by van der Marel et al. \markcite{vdM94}(1994), as well as with earlier
work by Kormendy \markcite{Kor88}(1988), both of which showed the dispersion
maximum coinciding with P2 to within the angular resolution. To further
complicate matters, the latter authors found a displacement of the center of
the rotation curve away from P2, in the direction {\it toward\/} P1.

An interesting effort to decompose the LOSVD in the P1--P2 region into
discrete kinematic components was undertaken by Gerssen, Kuijken, \& Merrifield
\markcite{GKM95}(1995) using long-slit spectra. They detected an asymmetric
tail extending to positive velocities. Modeling the
LOSVD as a sum of two Gaussians, they argued that the lower-amplitude,
higher-velocity component should be identified with P1. Gerssen et al.
did not go so far as to assert that P1 actually is a distinct physical
component. However, Emsellem \& Combes \markcite{EmC97}(1997), modeling the
TIGER data, have vigorously advocated a model in which P1 is a dense star
cluster on the verge of tidal disruption by the central black hole.

The mystery deepens still more when we look at the
population characteristics of the two peaks of brightness. As
shown by King, Stanford, \& Crane \markcite{KSC95}(1995), P2
increases in brightness at shorter wavelengths, whereas P1 has
the same color as the surrounding bulge light, from the far
ultraviolet to the near infrared. Lauer et al.
\markcite{Lau98}(1998), using post-repair \hst\ imaging, show
that P2 is bluer than its surroundings
throughout the optical range of wavelengths, and in addition
find that P2 has a resolvable core profile. The color properties
of both P1 and P2 are surprising.  The changing color of the
dynamical center P2 suggests the presence of some other source of
radiation in addition to the stars that make up the bulge.  Even more
perplexing, however, is the color of P1.  The interpretation of
such an object as a large globular cluster or a small dwarf elliptical galaxy
is tempting, but the colors militate against this.  Either a
globular cluster or a dwarf elliptical would be expected to have a
metallicity considerably below solar, whereas the color of P1 is, over a
large range of wavelength, a near-perfect match to the color of its
surroundings, which have a metallicity that is above solar.

Tremaine \markcite{Tre95}(1995) suggested an ingenious model that, in
principle, could simultaneously solve the population dilemma of P1 and
explain the nuclear kinematics. The essence of the model is that
P1 is not a self-bound object, but only a statistical accumulation of stars near
the apocenters of their orbits in an eccentric Keplerian disk surrounding
the black hole, which lives within P2. Tremaine was able to produce an
excellent fit to the Lauer et al.\ \markcite{Lau93}(1993)
photometry and Kormendy's \markcite{Kor88}(1988) kinematics. While the
question of how to form such a disk in a galactic nucleus is far from
trivial, it is at least possible that the configuration, once formed, could
be long lived, in contrast to models where P1 is a bound object on the
brink of destruction.

The high spatial resolving power of \hst\ has already led, as indicated
above, to a sharper knowledge of the spatial properties of the nuclear
region of M31. In this paper we investigate the kinematics with a
similarly improved resolution. Our presentation is organized as follows:
In Sec.\ \ref{s.observations} we describe the long-slit capabilities of the
\hst\ Faint Object Camera and our observational program. Issues of data
reduction, which are substantial, are discussed in Sec.\ \ref{s.reduction}.
Sec.\ \ref{s.kinematics} describes the methods used to extract the kinematic
profiles and presents the results. In Sec.\ \ref{s.discussion} we
compare the kinematics as revealed by the FOC with groundbased data and
with the published dynamical models. Finally, we summarize the main results
in Sec.\ \ref{s.summary}.

\section{Observations\label{s.observations}}

\subsection{The f/48 Spectrograph\label{s.spectrograph}}

Until the installation of STIS, the only long-slit spectrograph on board
\hst\ was in the Faint Object Camera.  In the planning of that
instrument, the members of its Instrument Science Team had realized that
the $f/48$ camera of the FOC could be given a spectrographic capability if
a grating were installed in such a position that temporary insertion of
a diverting mirror into the light beam sends the latter to the grating
before it reaches the cathode.  The diverting mirror and grating are
tilted by a small amount, so that what this optical mode images onto the
cathode is not the usual imaging aperture but rather a long slit that is
just to the side of the imaging aperture.

The spectrograph slit measures 0.063 $\times$ 12.5 arcsec.  Spectra are
recorded over the wavelength range 3600 to 5400 \AA ngstroms, on a field
that extends 1024 pixels in the dispersion direction and 512 pixels in
the spatial direction; of the latter, 446 pixels cover the slit length.
Each pixel is $50 \times 25 \microm$ in size (56 mas $\times$ 1.7 \AA).

The image of the slit is 2.3 pixels wide on the cathode, thus degrading
the spectroscopic resolution in return for an increased throughput.
(This width was chosen by IRK and PC, as members of the IST,
specifically with the present program in mind.)

\subsection{Planning of the Program\label{s.planning}}

This program (GTO-6255) was part of the Guaranteed Time Observations
(GTO) of IRK.  In the division of GTO orbits between various projects,
it was assigned 9 orbits, of which 7 were to be used for the actual
spectroscopic exposure.  Calculations of surface brightness and
instrumental sensitivity had indicated that this exposure length would
produce an adequate spectrum, at least in the brightest central region.
The position angle chosen for the slit was $42\arcdeg$, which would lay it
across P1 and P2.

It had been planned to carry the program out much earlier, but the $f/48$
camera failed in 1992 and was unusable for more than 2 years.  When it
unexpectedly came back to life in 1995, the program was reactivated and
carried out.  Unfortunately the revived $f/48$ had problems of
background noise; to monitor these we took a flatfield image (using an
internal light source) and a dark image (shutter closed) during each
occultation period.  As we shall see below, the background anomalies
varied with time in such a way that we were unable to remove them
completely. 

After beginning the reductions, we found that data for wavelength
calibration were lacking; in particular, we needed to determine the
dependence of the wavelength scale on position along the
slit. Furthermore, it quickly became apparent that it would be
impossible to extract reliable kinematic profiles without a stellar
spectral template of approximately the right spectral type, observed
with the same instrument. IRK and RIJ therefore prepared a calibration program
(CAL/FOC-6926) which secured a well-exposed spectrum of Jupiter, with
the planet covering nearly the entire length of the slit.  Because of
the danger of undispersed light of Jupiter coming through a direct path
that exists in the spectrograph, this exposure had to be made at the
limb of Jupiter's disk. Thanks largely to the absence of absorption
bands in the visible from the main constituents of the Jovian
atmosphere, the result was essentially a reflected solar (G2) spectrum.

\subsection{Acquisition\label{s.acquisition}}

M31 was observed during 10 orbits of \hst\ on 4--5 December 1995 UT.
The first orbit was devoted to the direct imaging of the target
that is required for every pointing of the spectrograph slit.  This
image is downlinked to the observer, who indicates where the slit
should be placed.  A pointing change is then made that should place the
slit correctly.  Unfortunately, however, the location of the slit relative
to the imaging format was
not well enough known to place it precisely across the centers of P1 and
P2.  We therefore took five exposures of 357 s each over the next two orbits,
displacing the slit
laterally by 0\secspt05 each time, and each of these was downlinked as
soon as it was complete.  We collapsed each of these images in
the dispersion direction, so as to provide a brightness profile, and we
used the intensities of these to judge where the center of M31 lay,
among the 5 positions.  We believe that our chosen position, at which
the slit was held for the next 7 orbits, is within 0\secspt05 of the
center of M31.

The total exposure in those 7 orbits was 19,097 s.

\section{Data Reduction\label{s.reduction}}

\subsection{Cleaning\label{s.cleaning}}

The raw FOC data contained a number of artifacts introduced by the
detector that hampered our original efforts at processing. These
included: remnants of the detector fly-back, burn-in lines on the
photo-cathode, and analog-to-digital converter glitches, to mention a few.
In addition to these, which are primarily high-spatial-frequency
perturbations, there were also two larger artifacts:\ an arc
extending over nearly a quadrant, and a diffuse flare near the center,
with a diameter of more than 200 pixels.
These were of course of lower spatial frequency, and varied with time. We
developed a simple procedure that was able to remove most of these
artifacts; and, most importantly, the process removed those
high-spatial-frequency features which would affect our analysis of the
absorption lines most severely.

We had eight dark frames available, taken in the occultation periods
before and after the data frames. The cleaning process made use of these
darks in the following manner: (1) The dark frames were summed directly
using the relevant IRAF task. (2) The resulting image was edited to
remove the most obvious high-frequency features. Image editing was
preferred over a low-pass filtering technique. (3) The edited image was
gaussian-smoothed using a
symmetric gaussian with $\sigma=5$. (4) The smoothed image was
subtracted from the original summed dark image. The resulting background
frame now had the high frequency artifacts emphasized. (5) This
background frame was scaled using an empirically determined scaling
factor, and subtracted from the individual data frames. The scaling
factors were determined individually for each data frame, since the
artifacts varied from frame to frame. The dark-subtracted images were
then further processed as described in the following section.

This rather simple procedure produced images that, although not perfect
to look at, gave considerably more reproducible results when
analyzed. It took several iterations on several versions of the data
before we realized that the first step in the process should be this
cleaning, before the geometrical correction.

\subsection{Flattening and Geometrical Corrections\label{s.corrections}}

The raw data from the $f/48$ camera consisted of a $512 \times 1024$ array of
pixels approximately $50 \times 25 \microm$ in size (the so-called
``zoomed'' pixel format).  The first step was to ``dezoom'' the image by
replacing each $50 \times 25 \microm$ pixel with two $25 \times 25 \microm$
pixels, each with half of the flux of the zoomed pixel.  This made the
resultant image a $1024 \times 1024$-pixel square.

The FOC suffers from geometric distortion as a result of the electric
and magnetic fields used to focus the electron beam that is produced
when a photoelectron is detected at the photocathode.  There is
additional distortion that arises from the spectrograph optics.  The
geometric correction process had to remove this distortion and deliver
an image in which lines of constant wavelength are horizontal
(wavelength independent of position $x$ along the slit) and points at a
given location along the slit trace a vertical line (spatial position
independent of $y$ coordinate in the dispersion direction).

In the raw dezoomed data, lines of constant wavelength were tilted by
approximately $7 \arcdeg$ to the horizontal, and lines of constant
position along the slit were approximately $8 \arcdeg$ to the vertical. 
The nonlinear nature of the geometric distortion made these tilts
functions of position on the detector.  The nonlinear distortion
could be removed by using the grid of reseau marks evaporated onto the
photocathode; these spots, approximately $75\microm$ in size and forming a
regular grid with spacing exactly $1.5\mm$, were easily visible in internal
flatfield images.

A description of the calibration of FOC spectrograph data is given in
Voit \markcite{Voit96}(1996) and in Jedrzejewski \& Voit
\markcite{JeV97}(1997). Spectra of the planetary nebula NGC 6543 were
used to regularize the emission lines (lines of constant wavelength),
and spectra of a star at two positions along the slit were used to map
the distortion in the spatial direction. The star was imaged both near
the center of the slit and $3\arcsec$ along the slit in the direction
away from the imaging format. The geometrical correction that made the
trace of the star in the first spectrum exactly vertical also rectified
the second spectrum to a good approximation. From this we deduced that
the distortion was not dependent on the position of the object on the
slit. The expense of these calibration exposures in telescope time
precluded a more careful mapping of the distortion.

Unfortunately, the geometric distortion in the FOC changes with time
in an unpredictable way, with typically a slight rotation of the field
of a few tenths of a degree over several orbits, and changes in the 
effective plate scale of a few tenths of a percent over the same period.
This means that the transformation used to correct the calibration data
was not applicable to the science data described here; the
time-dependent distortions are not removed. However, the correction for
the geometric distortions transforms the reseau marks to a well-defined set
of positions. Non-calibration FOC spectra can therefore be corrected by
transforming the data so that the reseau marks are transformed into the
positions of the marks in the optimally calibrated data.  Having taken
internal flatfields in the occultation periods between the M31 science
exposures, we were able to measure the coordinates of the reseau marks
that correspond in time with the science data and then generate custom
geometric correction reference files to remove the time-dependent component
of the distortion. Using this method gave geometrically corrected spectra
with only one resampling of the data, preserving most of the information on
the smallest scales.  It also allowed the re-transforming of the data to
take account of small single-orbit shifts of the spectra, as described
in Section \ref{s.registration}.  Thus the geometric transformations and
the positional shifts could be done in a single pass, and the only further
resampling needed was the conversion to a logarithmic wavelength scale,
which was done after wavelength calibration (Sec.\ \ref{s.calibration}).

Conventional flatfielding, such as is applied to direct images, was not
carried out.  First, no spectroscopic flatfield exists; second,
conventional flatfielding, as carried out for the FOC, affects only low
spatial frequencies that play no role in the work that we are doing here.

\subsection{Registration\label{s.registration}}

After correction for geometrical distortions, the single-orbit spectra
were still not positioned identically on the detector.  The
displacements must arise within the spectrograph rather than from a
shift in telescope pointing, as they exist both in the spatial ($x$)
direction and in the wavelength ($y$) direction.  We believe that the
errors originate from the removable spectrograph mirror, which had to be
removed during each occultation period for the flatfield and dark
exposures, and then replaced for the M31 spectrographic exposure.  We
believe that the mirror fails to seat exactly the same each time.

The relative offsets in the spatial ($x$) direction were computed by TSS
by summing the spectra in the $y$ direction over the central 800 lines
to obtain 1-dimensional spatial profiles, then cross-correlating these
against each other. The spatial offsets were also determined
independently by IRK in a similar way that differed only in detail; the
results were in agreement within a few tenths of a pixel.  The offsets
in the wavelength direction were calculated by TSS as part of the
wavelength calibration, as described below.

\subsection{Wavelength Calibration\label{s.calibration}}

The FOC spectrograph contains no internal arc lamps for wavelength
calibration; moreover, the instrument has been so little used that there
are no contemporaneous observations of emission-line sources
that could be used for the same purpose. Therefore, we used
a ground-based spectrum of M31, obtained by Ho, Filippenko, \& Sargent
\markcite{HFS93}(1993, hereafter HFS) with the Hale 5 m and kindly provided
to us by Luis Ho and Alex Filippenko, to calibrate both the M31 and
Jupiter spectra.
This differed from the straightforward approach, which would have been to
derive a wavelength scale from identified lines in the Jupiter spectrum
and assume that the same scale applied to M31. We chose to calibrate the
M31 spectrum independently as a consistency check. Below we describe the
method used to perform this calibration with the groundbased spectrum.
This method subsequently proved to be very robust in calibrating the
Jupiter spectrum, and superior to a ``traditional'' line-ID approach.
Thus both galaxy and template spectra are calibrated to the
HFS spectrum, which consequently defines the velocity zero point
for our derived kinematics.

The $2\arcsec$-wide slit (at position angle $77\arcdeg$) used by HFS 
projects, at our slit PA of $42\arcdeg$, into a $3\farcs 5$ length along
our slit, corresponding to 125 of our columns.  Thus 
from each of the individual $x$-shifted M31 frames we extracted a 125-column
sum centered on P2. Since
this included the entire well-exposed part of the spectrum, the final
calibration was not sensitive to the chosen width. Each resulting 1-D
spectrum was then matched to the HFS spectrum using a two-parameter
cross-correlation. That is, given the HFS spectrum $S(\lambda)$ and the
uncalibrated spectrum $G(y)$, we assumed a mapping from pixel number
to wavelength of the form $\lambda = a y + b$. We then computed
\beq\label{e.wavecal}
\xi(a,b) = \int S(\lambda) G\left({\lambda-b \over a}\right) d\lambda,
\eeq
and maximized with respect to $a$ and $b$. For the individual frames $a$
and $b$ were not well constrained separately, but a linear combination of
them was, and its slope was nearly the same for all frames.
We could therefore obtain the relative offsets of the spectra by
comparing the best-fit $b$ values at fixed $a$. We chose to do this
at $a=-1.7\ang$/pixel; for $-1.69 \leq a \leq -1.71$ the derived offsets
differed by $< 0.3$ pixel. (Negative $a$ means merely that the short-wavelength
end of the spectrum was at the high-pixel-number end of the detector.)

To reduce the effects of accumulated roundoff error in rebinning the
images, the complete set of geometrical corrections, including the $x$ and
$y$ offsets, was then re-done in a single pass. The seven resulting frames
were co-added, and $\xi(a,b)$ recomputed for the same 125-column sum. With
the higher $S/N$ of the co-added spectrum, $a$ was much better constrained,
the best-fit value being $-1.7012\ang$/pixel. All subsequent analysis was
performed on the wavelength calibrated co-added spectrum in de-zoomed
format. Even though the individual exposures contained no information
on scales smaller than the zoomed pixels, the time-variable geometric
distortions (Sec.\ \ref{s.corrections}) and image shifts
(Sec.\ \ref{s.registration}) produced an unintentional, though beneficial,
dithering on the detector, so that some of the higher resolution can
be recovered in the co-added image. Still, features smaller than two
de-zoomed pixels wide should be treated with some skepticism.

The three individual exposures of Jupiter were summed in the spatial
direction, avoiding reseau marks and cosmetic defects. The 1-dimensional
spectra were of sufficiently high $S/N$ that the residual $y$ shifts
could be determined by eye to within $0.1$ pixel. The shifted spectra
were then coadded. Wavelength calibration was slightly complicated by
the radial velocity difference between Jupiter and M31. In calculating
$\xi(a,b)$ the Jupiter spectrum was Doppler shifted (\ie\ logarithmically
in wavelength) to an assumed velocity
$v_{\rm J} - v_{\rm M31} = -300 \kms$ before integrating. Otherwise,
calibration was done as described above, with the result $a =
-1.7017\ang$/pixel.  There was no sign of a need for a higher-order term
in the wavelength solution.  The $0.03\%$ difference between this value
of $a$ and that obtained from the M31 data is insignificant. Owing to
the higher $S/N$ in the Jupiter spectrum, and since the larger value of
$a$ is still consistent with the M31 spectrum, we adopted this result
for both.

Because the calibrated Jupiter spectrum was blueshifted to the velocity of
the HFS spectrum, all velocities in this paper are measured relative to
the HFS mean, which is an average over an $8 {\,\rm arcsec}^2$ aperture
centered on the nucleus. Residual calibration errors may introduce a
systematic offset in the velocity scale; however, we believe any such
offset should be $<50\kms$. Mean errors in the $y$ registration
offsets of $\sim 0.3\,$pixel would generate a $\sim 35 \kms$ systematic error
in velocity dispersion, which would add in quadrature. Our derived
dispersions are substantially above $100\kms$, so this is not likely to be a
significant effect.

\section{Stellar Kinematics of the Double Nucleus\label{s.kinematics}}

\subsection{Methods \& Strategies\label{s.methods}}

\begin{figure}[t]
{\hfill\epsfxsize=2.4in\epsfbox{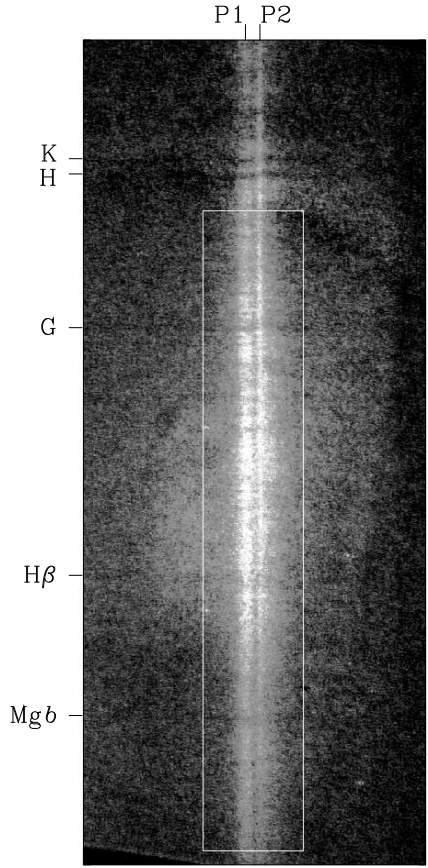}\hfill}
\caption{\footnotesize Final cleaned, geometrically corrected,
co-added spectrum of the M31 nucleus. The spectrum runs from $\sim 3670\ang$
at the top to $\sim 5500\ang$ at bottom. Bright vertical
streaks are P1 (left) and P2 (right). \ion{Ca}{2} H+K are easily
visible near the top; the G band, H$\beta$ and Mg $b$ are less
conspicuous. Note the residual smooth unsubtracted background and
the curved residual ``arc'' feature that contaminates H+K.  White box shows
the section of the spectrum used for the kinematic analysis.
\label{f.twodspectrum}}\end{figure}

The best-exposed section of the co-added M31 spectrum is shown
in Figure \ref{f.twodspectrum}. The figure covers the region from about
$3670\ang$ to $5500\ang$, and extends $6 \arcsec$ either side of P1 and
P2, which produce the parallel streaks down the middle of the image.
\ion{Ca}{2} H+K are easily visible near the top, and, less conspicuously,
the G band, H$\beta$, Mg $b$, and a number of weaker lines.
Also visible are residuals from the incomplete subtraction
of the arc and flare and imperfectly cleaned cosmetic defects. The
$S/N$ ratio reaches no higher than 14 (per pixel) at the centers of P1 and P2,
making the extraction of the stellar kinematics challenging. For
comparison, one generally aims for $S/N \gtrsim 20$ in ground-based 
absorption-line work to derive mean velocities and dispersions, and
$S/N > 30$ to obtain higher moments. Furthermore, in this case we could not
substantially increase $S/N$ by coarse spatial binning since the noise is
dominated by background systematics.

Nonetheless, even a casual inspection of H+K shows clear rotation in the
double nucleus. Lulled into false confidence by the Ca lines, we proceeded
with the kinematic analysis, only to meet with
disappointing---as well as borderline unphysical---results. After a
good deal of wasted effort we realized that \ion{Ca}{2} H+K were being
contaminated by residual background from the arc, and the correlation signal
corrupted by the disproportionate contribution of these two strong lines. 
The results improved dramatically when we discarded the Ca lines and
relied on the combined signal from the much more numerous weak lines in
the spectrum.

The region of the final image used in our subsequent analysis is marked by
the box in Figure \ref{f.twodspectrum}. One-dimensional spectra extracted
from this region were divided by continua fitted using high-order cubic
splines. They were then rebinned onto a logarithmic wavelength scale with
833 bins between $3984.39\ang$ and $5453.00\ang$, corresponding to a
velocity scale of $113.06 \kms {\rm pixel}^{-1}$. The same was done for
the Jupiter spectrum.

Because $S/N$ was far too low to allow nonparametric extraction of the LOSVD,
we assumed Gaussian broadening functions and recovered the mean
velocity $V$, dispersion $\sigma$, and normalization (``line strength'')
$\gamma$ using Statler's \markcite{Sta95}(1995) implementation of the
cross-correlation method. (We deal with the $h_3$ and $h_4$ Gauss-Hermite
coefficients in Section \ref{s.gausshermite}.) The cross-correlation
method exploits the following identity: if the galaxy spectrum $G$ is the
convolution $S \otimes B$ of the stellar template spectrum $S$ and the
broadening function $B$, then the galaxy-template cross-correlation function,
$X = G \circ S$, and the template autocorrelation function, $A = S \circ S$,
are related by $X = A \otimes B$. In our implementation, a region of
specified width around the primary cross-correlation peak is fitted with
a parameterized $B$ convolved with $A$, and the parameters are
manipulated to optimize the fit. For high $S/N$ data the results are not
very sensitive to the size of the fitting region; however, this becomes an
important issue for the present data.

Prior to cross-correlating, galaxy and template spectra were filtered in the
Fourier domain to remove low-frequency components. The filter was zero below a
threshold wavenumber $k_L$ (measured in inverse pixels), unity above $2k_L$,
and joined by a cosine taper in between. It is convenient to give the filter
width $W_T$ in Fourier-space pixels, in terms of which $k_L = W_T/833$.

\begin{figure}[t]{\hfill\epsfxsize=3.0in\epsfbox{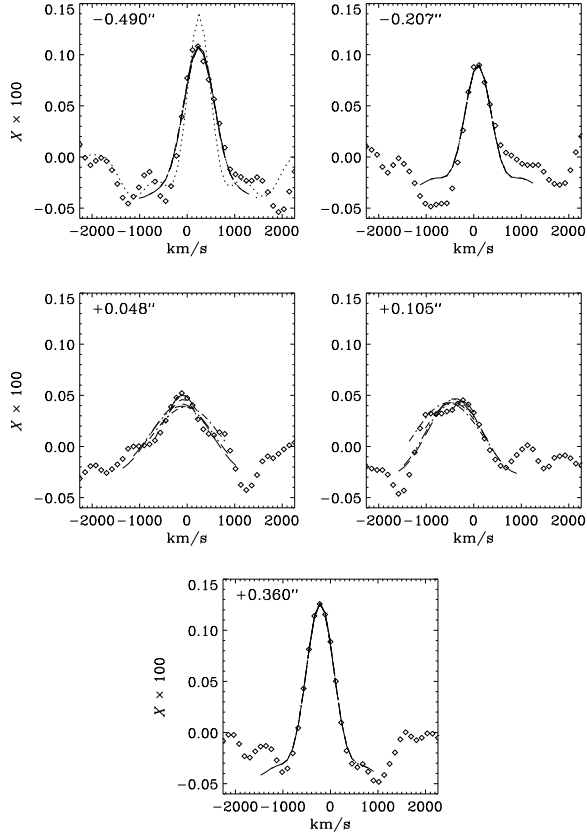}\hfill}
\caption{\footnotesize Primary peaks in the M31--Jupiter cross-correlation
functions ({\em points\/}) for galaxy spectra at five points along the slit,
computed with a filter width $W_T = 10$ (see Sec.\ \protect{\ref{s.methods}}).
Positions are given relative to P2. Overplotted
({\em smooth curves\/}) are the fitted functions obtained by convolving the
template autocorrelation function ({\em dotted line\/} in the first panel)
with the LOSVD. Different line styles show results of confining the fit
to windows within $\pm W_{\rm fit}= 2,3,5,7,9,11$ pixels of the maximum of $X$.
The results are sensitive to $W_{\rm fit}$ only in the region of the
dispersion spike ({\em middle two panels\/}); however, even the smallest
window gives values of $\sigma$ substantially higher than away from the spike.
\label{f.corrpeaks}}\end{figure}

The correlation peaks at five points along the slit, computed with
$W_T = 10$, are shown in Figure \ref{f.corrpeaks}. The point at
$-0\farcs 49$ is the center of P1. (All distances are measured with
respect to
the P2 brightness maximum.) The point at $+0\farcs 36$ is approximately
symmetrically placed on the other side of the rotation curve center;
comparison shows a clear velocity difference of $\sim 450 \kms$.
On top of the data are plotted the fitted
curves, $X_{\rm fit} = A \otimes B_{\rm fit}$, where the fits are confined
to windows within $\pm W_{\rm fit}$ pixels of the maximum of $X$. We
show the results for $W_{\rm fit} = 2,3,5,7,9,11$. At $-0\farcs 49$ and
$+0\farcs 36$ the results are very robust, as all the fitted curves fall
on top of each other. At $-0\farcs 207$, in the brightness minimum between
P1 and P2, $S/N$ has come down a bit and the wings of the correlation peak 
are less reliable; but again the inferred $V$ and $\sigma$ are still
insensitive to the fitting width $W_{\rm fit}$. This is typical of the
entire spectrum between $-1\farcs 0$ and $+0\farcs 4$, {\em except\/} for
a very small region about $0\farcs 25$ wide centered near P2. The
remaining two frames in Figure \ref{f.corrpeaks} show the correlation peaks
at two points near the center of this region. These points are separated by
one ``zoomed'' pixel width, approximately 1 FWHM of the PSF, so the data
are essentially independent. The correlation peaks have clearly broadened out
and come down in amplitude. The maximum amplitude is now comparable to that of
noise features,\footnote{Remember that errors in the cross-correlation
function are strongly correlated, so smooth-looking features can result from 
noise in the spectra.}
and it is difficult to tell what the real extent of the peak is. As a
result, the fitted $X_{\rm fit}$ depends sensitively on the width of the
fitting window.

To gain some control over this systematic uncertainty, we exploited the well
known covariance between velocity dispersion and line strength common to
all LOSVD
fitting methods. A normalized Gaussian has a central amplitude proportional
to $\sigma^{-1}$, and inspection of Figure \ref{f.corrpeaks} shows that the
amplitude of $X$ comes down in the exceptional region by about a factor of 2.
If there is no variation in the intrinsic line strength, this would indicate
about a factor 2 increase in dispersion. In the plotted examples we obtain
lower dispersion ($\sim 340 \kms$) for small $W_{\rm fit}$
since only the top part of the peak is fitted, but the fit also requires
a low $\gamma$ ($\sim 0.5$) to pull the amplitude down to match the data.
Similarly, wide $W_{\rm fit}$ gives large $\sigma$ and large $\gamma$.
Over the rest of the spectrum between $-1\farcs 0$ and $+0\farcs 4$ the
value of $\gamma$ is quite stable, with an average of $1.02$. We therefore
decided to constrain $\sigma$ by {\em assuming\/} that there is no significant
variation in $\gamma$ in this small region. We ran the correlation analysis
over a grid of filter widths ($W_T=8,10,12,15,18$) and fitting widths
($W_{\rm fit} = 2,3,5,7,9,11$), and took a weighted average of the
results. The weights for each point $i$ in the $(W_T,W_{\rm fit})$ grid
were set independently at each point along the slit, according to
$w_i = \exp\left[-(\gamma_i-1.02)^2/(2 s^2)\right]$. The width $s$ was set
to $0.1$, exponentially excluding combinations of $W_T$ and $W_{\rm fit}$
giving more than a $10\%$ variation in $\gamma$.

An additional complication to this strategy is the unsubtracted smooth
background still present in the final spectrum. This should enter as
essentially an additive constant, lowering the line strength by a
multiplicative factor and leaving the dispersion unchanged. We have not
attempted to correct for this effect explicitly since the background is
a relatively minor contribution in the P1--P2 region and the
model-dependent correction could tend to overemphasize regions of higher
dispersion.

Pushing out to larger radii proved to be very difficult, and we found that
even semi-objective strategies like the one just described behaved badly. 
We were forced simply to choose $W_{\rm fit}=2$ in order to keep the
dispersion consistently low enough to agree with existing groundbased
data. We took an unweighted mean of the results using $W_T=8,10,12,15$,
and obtained outer kinematic profiles that are at least consistent with
the expected behavior of $V$ and $\sigma$ around $2\arcsec$ from the
center. However, we do not have a great deal of quantitative confidence in
our results for this region, and we plot them with different symbols in
the figures below.

\subsection{Rotation and Dispersion Curves\label{s.results}}

\begin{figure}[t]{\hfill\epsfxsize=3.5in\epsfbox{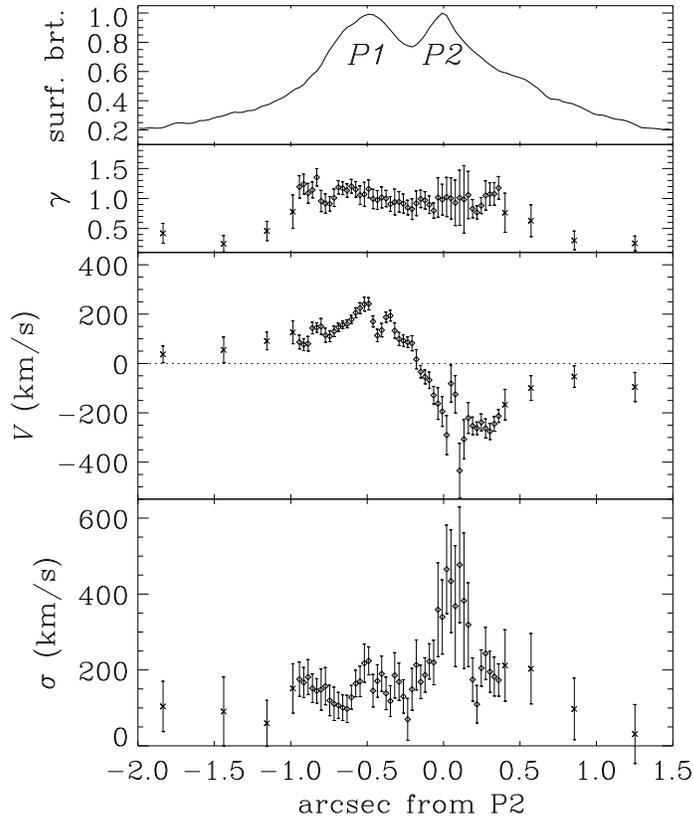}\hfill}
\caption{\footnotesize M31 nuclear rotation curve and
dispersion profile, at ``de-zoomed'' pixel resolution. {\em Top
panel\/}, approximate surface brightness profile along the slit.
{\em Second panel\/}, constrained line strength parameter $\gamma$.
(This is {\em not\/} an actual measurement of line strength; see Sec.\
\protect{\ref{s.methods}}). {\em Bottom two panels\/}, rotation curve
and dispersion profile. {\em
Diamonds\/} show the best data, to which $\gamma$-weighted averaging has
been applied. {\em Crosses\/} indicate radially-summed data for which no such
averaging was feasible and which are prone to large systematic errors.
\label{f.singlecols}}\end{figure}

Our results for the nuclear rotation curve and dispersion profile are
shown in Figure \ref{f.singlecols}. In the top panel, the
approximate surface brightness profile along the slit is shown for
orientation. This is merely the integral of the un-flux-calibrated
spectrum over wavelength.
We have subtracted a
constant background level of $0.25$ of the maximum to compensate for the
residual smooth background and to approximately fit the Lauer
et al. profile.
P1 and P2 appear here to be of comparable
brightness, even though P1 is substantially brighter in the $V$ image of Lauer
et al.\ \markcite{Lau93}(1993). This is attributable to color differences
between the two peaks (King et al.\ \markcite{KSC95}1995, Lauer et al.\
\markcite{Lau98}1998); P2 is actually 
brighter than P1 shortward of about $4300\ang$.
The second panel shows
the constrained line strength $\gamma$. Recall that we are forcing the $\gamma$
profile to be flat between $-0\farcs 95$ and $+0\farcs 4$, so this is
{\em not\/} a measurement of the line strength; it should be read as an
indicator of systematic errors. Note that at larger radii (crosses)
$\gamma$ falls, partly as a result of the narrow fitting window, and
partly because of the unsubtracted background.

The noteworthy features of the $V$ and $\sigma$ profiles,
in order of decreasing certainty, are:
\begin{enumerate}
\item The velocity zero crossing is between P1 and P2, $0\farcs 16 \pm
0\farcs 05$ from P2. We will refer to this point as the rotation center (RC).
\item The rotation curve through RC is resolved; the projected velocity
gradient at RC is approximately $300 \kms\pc^{-1}$ (for an assumed distance
of $725 \kpc$).
\item Except for small regions near P1 and P2, the rotation curve is
essentially symmetric about RC, with maximum amplitude 
$\sim 250 \kms$ reached about $0\farcs 34$ from RC.
\item The velocity dispersion is $\lesssim 250\kms$ everywhere except for
the small region, $0\farcs 25$ wide, centered $0\farcs 06 \pm 0\farcs 03$
past P2, which we call the dispersion spike.
\item The maximum $\sigma$ in the dispersion spike is $440 \pm 70 \kms$.
\item There may be a disturbance to the rotation curve at P1 which has the
signature of a marginally resolved local rotation in the same sense as the
general rotation of the nucleus.
\item There may be a slight elevation in $\sigma$ at P1.
\item There may be a large unresolved feature in the rotation curve
within the dispersion spike.
\end{enumerate}

\subsection{Further Tests\label{s.tests}}

The most striking feature in the kinematic profiles is the dispersion
spike. We performed several additional tests to verify that it is real.

\begin{figure}[t]{\hfill\epsfxsize=3.5in\epsfbox{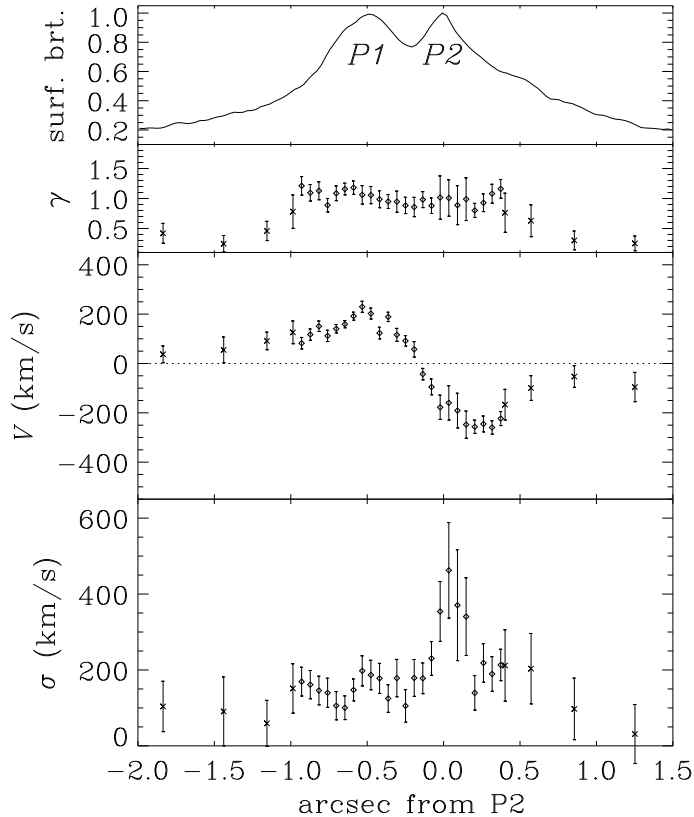}\hfill}
\caption{\footnotesize Same as Figure \protect{\ref{f.singlecols}},
but with the best data double-binned to the original ``zoomed'' pixel scale.
\label{f.doublecols}}\end{figure}

First, we summed pairs of columns in the de-zoomed image
to increase $S/N$ and repeated the kinematic analysis at half the
spatial resolution (that is, going back to the original zoomed pixel
size). The double-binned kinematic profiles are shown in
Figure \ref{f.doublecols}. All of the major features of Figure
\ref{f.singlecols} are reproduced, including the dispersion spike. Marginal
features that are unresolved at single-column resolution are smoothed over
by the coarser binning. 

Second, we double-binned the de-zoomed single-column spectra to half the
{\em spectral\/} resolution and again re-did the analysis. This produced
essentially the same cross-correlation peaks as the examples shown in
Figure \ref{f.corrpeaks}, but sampled at twice the original spacing.
The broadening of the peaks in the region of the dispersion spike was not
affected.

Third, to test whether the spike could be the result of a locally
contaminated region of the image (like the contamination of Ca H+K
discussed above), we broke the spectrum into red and blue halves and
again repeated the analysis. To preserve a workable $S/N$
ratio it was necessary to work with the double-summed columns. Once
again the kinematic profiles were qualitatively unchanged.

Finally, concerned whether an error in registration or some other unrecognized
single-frame problem could be producing a spurious feature, we
constructed 7 co-adds of 6 of the frames, excluding each one in turn.
As before, the kinematic profiles were not significantly affected.

We should emphasize that, although the amplitude of the spike depends on
the cross-correlation fitting width $W_{\rm fit}$, the spike is still present
even when only the narrowest, most sharply curved section of the correlation
peak is fitted. And while it is possible that our results for P2 could be
affected by its unresolved center behaving effectively as a point source in
our relatively wide slit, point-source contamination would tend to make
the line profile narrower, not broader.

As a result of these tests, we believe that the dispersion spike is a
real feature, and that at least items 1--5 in our list above accurately
describe the stellar kinematics of the nucleus.

\subsection{Gauss-Hermite Terms\label{s.gausshermite}}

The higher moments of the LOSVD are potentially important constraints on
models of the nuclear dynamics. Unfortunately, we have insufficient
signal to be able to recover these quantities at anything but the coarsest
of resolutions. We broke the best-exposed region into two $0\farcs 53$
(19-column) bins on either side of the RC and extracted the LOSVD
using a Gauss-Hermite expansion for the
broadening function (van der Marel \& Franx \markcite{vdMF93}1993,
Statler \markcite{Sta95}1995), truncated at fourth order. The $h_3$ and
$h_4$ coefficients are related, respectively, to the skewness and kurtosis
of the LOSVD. The results again depend on the filter and fitting-width
parameters. We choose $W_{\rm fit}$ to encompass the region between the
first maxima on either side of the primary correlation peak, and adopt
$W_T=12$ simply on the grounds
that this combination gives $\gamma \approx 1$. We obtain, for the P1
side, $h_3 = 0.21 \pm 0.35$, $h_4 = 0.08 \pm 0.50$, and for the P2 side,
$h_3 = -0.08 \pm 0.07$, $h_4 = 0.11 \pm 0.11$ (internal errors only).

\begin{figure}[t]{\hfill\epsfxsize=3in\epsfbox{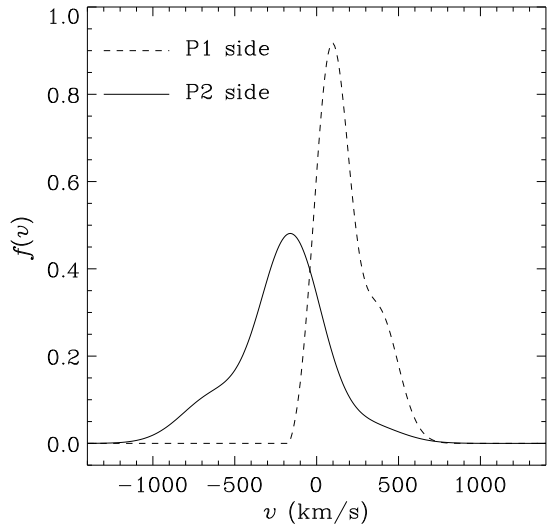}\hfill}
\caption{\footnotesize LOSVDs reconstructed from the derived $V$,
$\sigma$, $h_3$, and $h_4$ parameters for spectra summed over two
wide radial bins on either side of the rotation center. The wide wings
on the P2 side are consistent with the shapes of the correlation
peaks in the dispersion spike (Fig. \protect{\ref{f.corrpeaks}}). The
$h_3$ values, nonzero at marginal significance, suggest a
longer tail to the velocity distribution in the direction of rotation,
opposite to sense of skewness generally seen.
\label{f.losvd}}\end{figure}

The reconstructed LOSVDs for the two sides of the rotation center are
shown in figure \ref{f.losvd}. There is an indication of very wide wings
on the P2 side, consistent with the visual impression of the correlation
peaks in the dispersion spike. The $h_3$ values, while differing from
zero at only marginal statistical significance, are consistently of
opposite sign for all choices of $W_T$ and  $W_{\rm fit}$. Note, however,
that the longer tail of the distribution is in the direction of rotation,
which is opposite to sense of skewness generally seen.
This disagrees with the ground-based results
of van der Marel et al.\ \markcite{vdM}(1994), where $V$ and $h_3$ are
found to be of opposite sign, as usual; but whether their resolution
would have been sufficient to resolve a change of sign of $h_3$ in the
inner half arcsecond is not clear. We admit that this is statistically a very
weak result and eagerly await confirmation by STIS.

\section{Discussion\label{s.discussion}}

\subsection{Comparison with Groundbased Observations\label{s.groundbased}}

Several recent groundbased studies have focused on the rotation and
dispersion of M31's double nucleus. None has been able to resolve the
central rotation curve, and none has reported a velocity dispersion spike
of the magnitude we see
near the position of P2. In this section we address whether our results
(Figures \ref{f.singlecols} and \ref{f.doublecols}) are consistent
with the lower-resolution groundbased data.

\begin{figure}[t]{\hfill\epsfxsize=3.5in\epsfbox{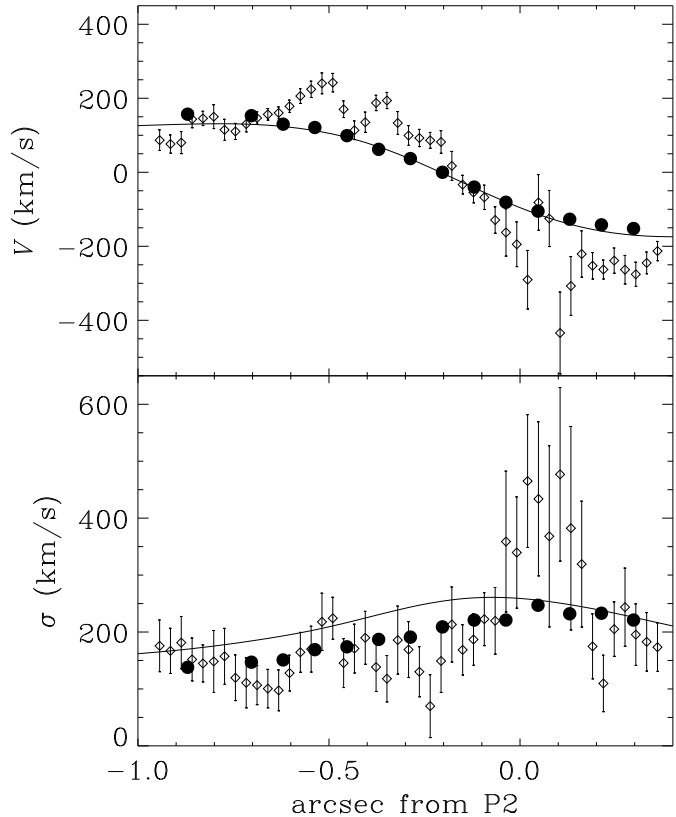}\hfill}
\caption{\footnotesize Comparison of the FOC kinematic profiles ({\em
diamonds\/}) with groundbased results from CFHT/SIS (Kormendy \& Bender
1999) ({\em circles\/}). {\em Smooth curves\/}
show our results convolved to CFHT resolution.
\label{f.kormendy}}\end{figure}

The highest-resolution groundbased work to date is that of Kormendy \&
Bender \markcite{KoB99}(1999, in preparation; hereafter KB), some results
of which are presented by Kormendy \& Richstone \markcite{KoR95}(1995).
KB used the Canada-France-Hawaii Telescope Subarcsecond Imaging Spectrograph
with a $0\farcs353$-wide slit, in $0\farcs63$ seeing (FWHM).
The kinematic results from their non-bulge-subtracted
spectra (read directly from Figure 7 of Kormendy \& Richstone
\markcite{KoR95}1995) are plotted as the filled circles in Figure
\ref{f.kormendy}. Our results are plotted as diamonds, and the plot is
confined to the region of our best data.

To make a valid comparison between the two data sets we must convolve our
data to CFHT resolution. This requires a two-dimensional convolution due
to the presence of structure on scales smaller than the SIS slit. Since
our slit is less than one-fifth as wide, we must make some assumptions for
the structure of the nucleus in the perpendicular direction.
We proceed as follows: let the coordinate on the sky in the direction
of the slit be $x$ and the perpendicular coordinate be $y$. First we build
a smooth 2-D model for the surface brightness distribution, $\mu(x,y)$,
that merges our 1-D spatial profile
with the Lauer et al.\ \markcite{Lau93}(1993) $V$-band image. The
model reproduces the Lauer et al.\ isophotes to within $\sim 0.2$
magnitude, and as the results are not sensitive to the details of the model,
there is nothing to be gained from a more exact fit. We extrapolate our
$V$ and $\sigma$ profiles to large $x$ (beyond the edge of the plot in
Figure \ref{f.kormendy}) by tacking the KB profiles onto the end of our data.
Since features beyond a few tenths of an arcsecond are resolved by KB, this
makes little difference to the convolved profiles in the inner region, and
eliminates any influence of our very uncertain results for larger radii.
We extrapolate the mean velocity field in the $y$ direction
by assuming no dependence on $y$; in other words, equal-velocity
contours are straight lines perpendicular to the slit. We do the same
for the dispersion field, except for the dispersion spike, which we assume
to be round, $\sigma$ falling with $y$ to
an ambient value of $180 \kms$. We then convolve the
three velocity moments, $\mu$, $\mu\langle v \rangle$, and $\mu \langle v^2
\rangle$, with an analytic fit to the PSF provided by John Kormendy,
and, at each $x$, integrate over a $0\farcs353$ wide boxcar window in $y$.
Finally, we recover the mean $V$ and dispersion $\sigma$ from the convolved
moments.

The resulting rotation curve, convolved to CFHT resolution, is shown as
the smooth curve in the top panel of Figure
\ref{f.kormendy}. (Because the spatial origin in the KB data is defined to
be the center of the rotation curve, we have shifted both KB profiles to
match the velocity zero point of our convolved rotation curve.) The velocity
gradient near the RC is reproduced extremely well; and while there is some
disagreement $\gtrsim 0\farcs5$ from P2, where our
curve is $\sim 30 \kms$ higher than KB's on the P1 side and $\sim 20
\kms$ lower on the anti-P1 side, this is consistent with the likely
level of systematic error in our velocity zero point. In our judgment,
the FOC rotation curve is completely consistent with the KB data.

Our convolved dispersion profile is plotted as the smooth curve in the
bottom panel of Figure \ref{f.kormendy}. Notice that at CFHT resolution
the dispersion spike all but disappears; the maximum dispersion in our
convolved profile is only about 5\% higher than KB's peak value, and
occurs only $0\farcs12$ from KB's peak, which is consistent with zero at
their resolution. The only significant disagreement in the dispersion
profiles is that ours is systematically high by about $50 \kms$ on the P1
side of P2. This is {\em not\/} attributable
to the dispersion spike. The convolved profile does ride above most of our
unconvolved data points, but not because of the spike, rather because of
unresolved rotation. Removing the spike entirely from our data and
repeating the convolution alters the profile by no more than $20\kms$, and
only in the vicinity of the spike; it makes no difference in the region
where the discrepancy is largest. The dispersion spike
would not have been detected by KB, because, at a characteristic size
$\sigma_{\rm spike} \approx 0\farcs1$, it is substantially smaller than
KB's $\sigma_\star = 0\farcs27$ seeing disk.

Our convolved profile could be reconciled with KB's if we were
systematically overestimating the dispersion by $\sim 50 \kms$. We suspect
that this may be the case, but unfortunately we have not been able to
determine why. We have tested the
cross-correlation code against artificial galaxy spectra made by
broadening the Jupiter template, and find no tendency to overestimate
$\sigma$ unless $\sigma \lesssim 40 \kms$.
We have repeated these tests
with a constant background level added to the artificial spectra, to
simulate the smooth unsubtracted background in the final M31 spectrum,
with the same result. It is possible that template
mismatch could induce a systematic error of this magnitude, but, lacking
other stellar spectra observed with the same instrument, we are unable
to assess this effect quantitatively.

An interesting decomposition of the nuclear LOSVD is presented by Gerssen,
Kuijken, \& Merrifield \markcite{GKM95}(1995), who obtained a spectrum
with the Multiple Mirror Telescope in $\sim 1\arcsec$ seeing, using a
$1\farcs25$ slit along PA $148\arcdeg$, or about $106\arcdeg$ from the
P1--P2 line. Within $\sim 1\arcsec$ of the brightest point along their slit,
they detected an asymmetric positive-velocity tail to the LOSVD. Using a
two-Gaussian decomposition, they interpret this tail as a discrete component
with a mean velocity of $+160\kms$ (relative to systemic) and a dispersion
$<100\kms$, and identify it with P1. In our case, we do not see a clear
signature of P1 as a separate kinematic component. Rather, P1 appears as
a brightness enhancement along the generally smooth $V$ and $\sigma$ profiles. 
Nonetheless, Gerssen et al.'s basic result of an asymmetric LOSVD can be
consistent with ours, depending on the centering of their slit.
Merrifield (1998, private communication) reports that the slit was centered
visually on the brightest point in the nucleus as seen on the slit-jaw TV.
This implies that the slit center was probably near P1, though, given
the slit width and seeing, an error of a couple of tenths of an
arcsecond is not unlikely. Nominally, their slit would extend
from about $-1\farcs1$ to $+0\farcs1$ on our scale, or displaced from
the RC toward the positive-velocity side. A centering error of $0\farcs25$ in
the direction away from P2 would put the RC at the edge of their slit,
and positive velocities would dominate. Contrary to Gerssen et al.'s
later comment, the LOSVD asymmetry does not rule out Tremaine's
\markcite{Tre95}(1995) eccentric disk model, since their slit would not
have covered the entire disk, and the negative-velocity side of their
observed LOSVD would be very sensitive to centering errors of this
magnitude.

Our results are also broadly consistent with the nuclear
major- and minor-axis $V$ and $\sigma$ profiles obtained by van der Marel
et al. \markcite{vdM94}(1994) using the William Herschel Telescope ISIS
spectrograph and the MMT Red Channel. Their major-axis rotation curve,
obtained in $\sim 1\arcsec$ seeing,
crosses zero $\sim 0\farcs2$ from the brightness maximum on their
slit (which we identify as P1) in the direction towards P2. On the nuclear
minor axis, on a slit also running through P1, they find $V \approx 50\kms$
at the brightest point. If their $0\farcs8$ slit were centered exactly
on P1, we might have expected, based on the FOC rotation curve, a somewhat
higher velocity; but seeing may have shifted the apparent brightness peak,
and consequently their slit, slightly toward the RC. Their major-axis
dispersion profile peaks within $0\farcs1$ of P2, as does our convolved
profile in Figure \ref{f.kormendy}.

Observations made at the CFHT with the TIGER integral-field spectrograph
are reported by Bacon et al.\ \markcite{Bac94}(1994). Their spatial
resolution was not quite as good as that of KB, but their data cover
a two-dimensional field on the sky, roughly $10 \arcsec \times 8 \arcsec$.
The general behavior of the TIGER kinematic fields in the direction
perpendicular to our slit supports our above assumption of
little variation in $V$ and $\sigma$ over the width of KB's aperture.
The TIGER rotation curve peaks roughly $0\farcs7$ from the RC at an
amplitude of $120\kms$, which is entirely consistent with the FOC rotation
curve convolved to a resolution slightly broader than in Figure
\ref{f.kormendy}. Puzzlingly, however, the TIGER dispersion field peaks
at $+0\farcs73$ from P2,\footnote{Actually Bacon et al.\ set their origin
at what they call the ``nucleus photometric center,'' which, from close
inspection of their Figures 13 and 26, we identify with P2.}
in the anti-P1 direction. We cannot confirm this
result with the FOC data. While our $\sigma$ values at $+0\farcs7$ are 
not especially trustworthy, we do see the dispersion falling steeply
from the spike out to $0\farcs4$, and is is difficult to see how the
apparent peak dispersion could be shifted that far if the true peak were
at P2. We have no explanation for this discrepancy. We can remark,
however, that Bacon et al.\ find the TIGER rotation center also on the
anti-P1 side of P2, which the FOC data contradict.

\subsection{Comparison with Published Models\label{s.models}}

There have been many efforts to model the central kinematics of M31,
mainly with the goal of estimating the mass of the likely black hole
(see Hodge \markcite{Hod92}1992 and Kormendy \& Richstone \markcite{KoR95}1995
for reviews). However, only two published models to date deal
quantitatively with the asymmetric structure of the nucleus, and
we confine ourselves to a comparison with these models.

\begin{figure}[t]{\hfill\epsfxsize=3.5in\epsfbox{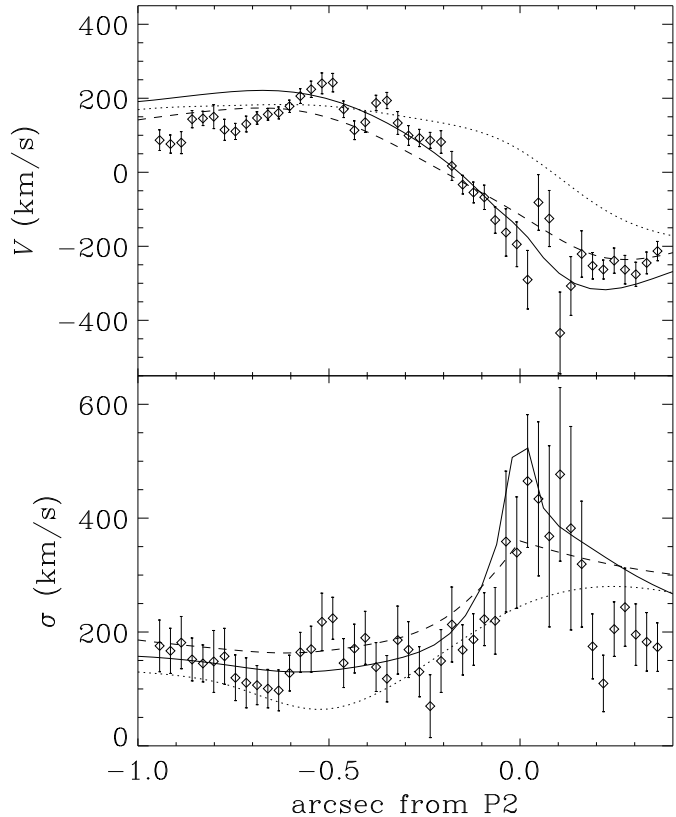}\hfill}
\caption{\footnotesize Comparison of the FOC kinematic profiles ({\em
diamonds\/}) with dynamical models at \hst\ resolution. {\em Dotted line\/},
sinking cluster model of Emsellem \& Combes
(1997). {\em Dashed line\/}, eccentric disk
model of Tremaine (1995), scaled up by 20\%
in velocity. {\em Solid line\/}, modified Tremaine-like model (see Sec.\
\protect{\ref{s.models}}).
\label{f.models}}\end{figure}

Emsellem \& Combes \markcite{EmC97}(1997, hereafter EC) consider P1 to be a
star cluster orbiting P2, the true center of the nucleus harboring the
black hole. The cluster must be of sufficient density that it largely avoids
tidal disruption up to the present time; but as a consequence dynamical
friction rapidly brings the cluster in to a radius where it {\em is\/}
disrupted. EC estimate that the present configuration will have a lifetime
substantially under a million years, but also argue that nuclear ingestion
of globular clusters should be a common event. EC present $N$-body
simulations of a system consisting of a static bulge, an axisymmetric
nuclear disk, a central dark mass of $9.4\times 10^7 \msun$, and an
orbiting cluster. They focus
primarily on comparison with the TIGER data, but plot kinematic profiles
for their best model at a resolution of $0\farcs06 \times 0\farcs06$, which
is a fairly close match to our $0\farcs063$ slit width and $0\farcs056$ zoomed
pixel scale. We plot these profiles, read directly from EC's Figure 15,
as the dotted lines in Figure \ref{f.models}, compared with our
results. Possibly due to the desire to fit the TIGER data,
their rotation curve crosses zero at about $+0\farcs1$ on the anti-P1 side
of P2, which is not consistent with the FOC data. The velocity dispersion
in the EC model is particularly low in P1, since it is assumed to be a
bound cluster of only a few million $\msun$. We do not detect a low $\sigma$
in P1, although we may have systematic errors in our dispersions, and even
with higher $S/N$ we would probably have difficulty measuring $\sigma \approx
60\kms$ as EC predict. EC's dispersion reaches a maximum of $290\kms$
$0\farcs25$ from P2, which is marginally consistent with, though qualitatively
different from, our dispersion spike. Since EC did not do a systematic study
of the parameter space, it is difficult to say whether their model could
be adjusted to fit the dispersion spike and the rotation curve. However,
the $\sim 200\kms$ dispersion we measure in P1, if confirmed by STIS,
would argue strongly against P1 being a bound cluster,
consistent with the argument of King et al. \markcite{KSC95}(1995)
that the P1 colors indicate a metallicity substantially above that of
typical globulars.

If P1 is not a bound object, it may still be produced by a statistical
accumulation of stars, as in the model of Tremaine \markcite{Tre95}(1995).
In this model P2 represents the stellar cusp associated with the black hole,
around which orbits an eccentric Keplerian disk. The line of apsides in the
disk is presumed straight or nearly so, and the brightness enhancement at P1
is caused by the accumulation of stars lingering near the apocenters of their
orbits. In this picture P1 is a purely kinematic $m=1$ density wave with zero
pattern speed. It is necessary that the eccentricity of the disk decrease
outwards, and also that its surface density decrease interior to P1,
in order to produce a ``blob'' in projection rather than simply an extension
of P2 or a ring around it. For the correct viewing geometry, the rapid
speeds of disk stars near pericenter place P2 at negative velocity
in projection, shifting the RC toward P1, as we observe. Tremaine plots
$V$ and $\sigma$ profiles for a model that fits the Lauer
et al.\ \markcite{Lau93}(1993) photometry and groundbased kinematics from
Kormendy \markcite{Kor88}(1988), both at groundbased and ``perfect''
resolution. We show the latter result as the dashed lines in Figure
\ref{f.models}. We have scaled up the original model by 20\% in velocity,
or $9.5\%$ in black-hole mass, to improve the
fit. The agreement in the rotation curve is admirable, though
the central slope is somewhat underestimated. The dispersion profile is
consistent with our data except in the dispersion spike.

We have computed a number of Tremaine-like models to test whether the
fit can be improved by minor tweaking of the parameters. An improved
model is shown as the solid line in Figure \ref{f.models}. This model
uses the same black-hole mass ($7.5\times 10^7 \msun$) as Tremaine's
original model. We emphasize, however, that in neither case has
the mass been treated as a free parameter; this discussion is {\em not\/} 
a measurement
of the mass of the central object, only an illustration of the
flexibility of the eccentric-disk picture. Our improved model has a slightly
higher eccentricity and a steeper eccentricity gradient.
In order to reproduce the dispersion spike and the rotation gradient at RC
we have had to rotate the line of apsides so that it aligns with the
line of nodes, with the result that the model does not reproduce
the PA difference between
the P1--P2 line and the nuclear major axis. It turns out to be rather
difficult to reproduce simultaneously the steepness of the rotation
curve, the location of the RC, and the overall photometric symmetry
of the nucleus about P2 beyond the distance of P1. Higher eccentricity
can sharpen the rotation curve but damages the photometric symmetry. The
narrowness of the dispersion spike poses another problem. The line-of-sight
dispersion can be kept low on the P1 side of the spike by the
contribution of the cold disk component, which, on that side, rotates slowly
along with the underlying inner bulge. On the anti-P1 side the disk is a
minor contributor to the projected light, and furthermore is well
separated in radial velocity from the bulge stars. As a result it is
extremely difficult to reproduce the dispersion gradient on the anti-P1 side
of the spike with a Tremaine-like model. It is conceivable that introducing
rotation in the central part of the bulge component---which in the Tremaine
model is non-rotating---could reduce the problem
a bit. Alternatively, a very steep
eccentricity gradient inward toward the inner edge of the disk could help,
but this poses technical difficulties for the models as presently formulated.

The value of demanding a more exact fit from what is a {\em very\/}
approximate model is dubious, however.
Tremaine's model, as published, contains many simplifying assumptions.
The disk is assumed infinitely thin, and approximated by thin
``ringlets'', broadened by a simple convolution in the plane of the sky.
Velocity dispersion in the disk is not included in a realistic way.
The self-gravity of the disk and the potential of the bulge are both ignored.
In our models we have corrected only the decomposition into ringlets,
adopting a continuous mass distribution. None of these approximations is a
failing {\it per se\/}, since at groundbased resolution a rigorously careful
treatment is not warranted. However, we are very near, if not at, the
point where dynamical models of the M31 nucleus will require self-consistency
in some measure to address more than just the gross characteristics of the
data. Detailed modeling is beyond the scope of this paper, but we hope
to report on further modeling efforts in a future publication.

\section{Summary\label{s.summary}}

We have used the $f/48$ long-slit spectrograph of the \hst\ Faint Object
Camera to observe the double nucleus of M31 in the spectral region from
$3670\ang$ to $5500\ang$. We acquired a 19,000-sec.\ exposure with the slit
oriented along the P1--P2 line. A calibration spectrum of Jupiter was
used as a spectral template in a cross-correlation algorithm to extract
the profiles of the projected mean velocity $V$ and velocity dispersion
$\sigma$. Though the analysis was complicated by detector backgrounds,
geometrical distortions, and low signal-to-noise ratio, we believe the
data convincingly show
\begin{enumerate}
\item a resolved rotation curve, with maximum
amplitude $\sim 250 \kms$ and center (zero point) on the P1 side of P2; and
\item a generally low ($\lesssim 250 \kms$) dispersion everywhere except for
a narrow ``dispersion spike'', centered very close to P2, where $\sigma$
peaks at $440 \pm 70 \kms$.
\end{enumerate}
At lower confidence, there may be unresolved
local disturbances to the rotation curve associated with P1 and P2, and
a slight elevation in $\sigma$ at P1. And at very low statistical
significance we find a weak signature of LOSVD asymmetry that is opposite
to that usually encountered, in the sense that the longer tail of the
distribution is in the direction of rotation.

Our results are generally consistent with previous groundbased work. In
particular, when we convolve our data to match the resolution of
Kormendy \& Bender's \markcite{KoB99}(1999) CFHT/SIS data, we find an
excellent agreement in the rotation curve, and a 20\% disagreement
in the dispersion profile that cannot be traced to the dispersion spike.
Our $V$ and $\sigma$ profiles are consistent with the data of Gerssen
et al.\ \markcite{GKM95}(1995) and van der Marel et al. \markcite{vdM94}(1994),
although the latter authors do not detect the same sense of LOSVD asymmetry
that we see. Our data are {\em not\/} consistent with certain details of the
CFHT/TIGER data of Bacon et al. \markcite{Bac94}(1994), specifically the
locations, relative to P2, of the rotation center and the maximum dispersion.

Finally, we have assessed the ability of two published models
to fit the FOC data. The sinking star cluster model of Emsellem \&
Combes \markcite{EmC97}(1997) does not reproduce either the rotation
curve or the dispersion profile. In particular, it places the
center of the rotation curve on the wrong side of P2, and predicts that P1,
being a bound, low-mass object, should have $\sigma \approx 60 \kms$,
whereas we measure $\sigma \sim 170\kms$. The eccentric disk model of
Tremaine \markcite{Tre95}(1995) comes closer to matching the FOC
profiles, if the central black hole mass is scaled up by 10\%. A
significant improvement to the kinematic fit can be realized by tweaking the
parameters, at some sacrifice to the photometric fit. The models,
however, are presently limited by simplifying assumptions that are less than
realistic. Quantitative dynamical modeling of the
M31 double nucleus will require models of significantly greater realism
and sophistication, especially when data from STIS become available.

\acknowledgments

TSS acknowledges support from NASA Astrophysical Theory Program Grant
NAG5-3050 and NSF CAREER grant AST-9703036, IRK from NASA Grant
NAG5-1607, and PC from NASA Grant NAS5-27760.  We thank John Kormendy for
providing details of the CFHT/SIS observations in advance of
publication, and Mike Merrifield and Marc Davis for useful information
and helpful suggestions.

\section*{Appendix: Kinematic Data}

Tables \ref{t.singlecols} and \ref{t.doublecols} present the derived
kinematic profiles in the region between $-1\arcsec$ and $+0\farcs4$ from P2.
Table \ref{t.singlecols} gives the results of the analysis on single-column
spectra from the de-zoomed images (Fig.\ \ref{f.singlecols}); some
information is present on these scales because of the orbit-to-orbit
positional shifts of the spectrum on the detector. Table \ref{t.doublecols}
gives the results for columns summed in pairs, corresponding to the original
zoomed pixel format (Fig.\ \ref{f.doublecols}).
Uncertainties include the systematic contribution from the choice of
filter and fitting parameters in the cross-correlation code, but not from
template mismatch or background corrections.

\begin{deluxetable}{ccccc}
\tablewidth{0pc}
\tablecaption{M31 Nuclear Kinematics --- De-Zoomed Pixel Resolution
\label{t.singlecols}}
\tablehead{\colhead{$R(\arcsec)$} & \colhead{$V$} &
\colhead{$\pm$} & \colhead{$\sigma$} &
\colhead{$\pm$}}
\startdata
$-0.943$ & $  86.8$ & $ 27.9$ & $175.7$ & $ 45.4$ \nl
$-0.915$ & $  76.3$ & $ 24.4$ & $166.9$ & $ 39.7$ \nl
$-0.887$ & $  80.1$ & $ 29.8$ & $181.4$ & $ 45.9$ \nl
$-0.858$ & $ 143.5$ & $ 23.3$ & $151.8$ & $ 37.7$ \nl
$-0.830$ & $ 145.6$ & $ 19.3$ & $144.8$ & $ 32.5$ \nl
$-0.802$ & $ 150.2$ & $ 32.2$ & $148.3$ & $ 54.4$ \nl
$-0.773$ & $ 114.8$ & $ 28.5$ & $157.3$ & $ 49.2$ \nl
$-0.745$ & $ 110.4$ & $ 21.8$ & $119.6$ & $ 40.1$ \nl
$-0.717$ & $ 130.4$ & $ 21.3$ & $110.8$ & $ 44.0$ \nl
$-0.688$ & $ 146.6$ & $ 16.8$ & $106.7$ & $ 34.4$ \nl
$-0.660$ & $ 156.5$ & $ 15.7$ & $100.6$ & $ 33.8$ \nl
$-0.632$ & $ 160.4$ & $ 16.1$ & $ 97.4$ & $ 35.9$ \nl
$-0.603$ & $ 178.4$ & $ 16.6$ & $127.9$ & $ 31.5$ \nl
$-0.575$ & $ 206.0$ & $ 19.9$ & $164.4$ & $ 35.0$ \nl
$-0.547$ & $ 224.3$ & $ 22.0$ & $169.9$ & $ 40.3$ \nl
$-0.518$ & $ 240.3$ & $ 28.2$ & $217.9$ & $ 50.4$ \nl
$-0.490$ & $ 242.3$ & $ 24.9$ & $224.2$ & $ 36.9$ \nl
$-0.462$ & $ 170.2$ & $ 22.4$ & $145.4$ & $ 43.0$ \nl
$-0.433$ & $ 113.6$ & $ 25.2$ & $170.9$ & $ 42.9$ \nl
$-0.405$ & $ 135.2$ & $ 27.4$ & $189.8$ & $ 46.5$ \nl
$-0.377$ & $ 187.4$ & $ 20.7$ & $138.5$ & $ 43.0$ \nl
$-0.348$ & $ 193.9$ & $ 21.9$ & $118.0$ & $ 40.7$ \nl
$-0.320$ & $ 133.2$ & $ 30.9$ & $185.8$ & $ 59.7$ \nl
$-0.292$ & $  99.4$ & $ 26.7$ & $169.0$ & $ 49.0$ \nl
$-0.263$ & $  92.7$ & $ 23.3$ & $130.2$ & $ 44.0$ \nl
$-0.235$ & $  86.3$ & $ 21.4$ & $ 69.8$ & $ 55.3$ \nl
$-0.207$ & $  81.9$ & $ 30.6$ & $149.1$ & $ 55.2$ \nl
$-0.178$ & $  17.2$ & $ 39.1$ & $213.2$ & $ 65.9$ \nl
$-0.150$ & $ -33.3$ & $ 25.2$ & $168.6$ & $ 44.1$ \nl
$-0.122$ & $ -54.6$ & $ 28.1$ & $186.9$ & $ 45.2$ \nl
$-0.093$ & $ -67.7$ & $ 32.9$ & $222.6$ & $ 46.5$ \nl
$-0.065$ & $-129.0$ & $ 35.7$ & $219.7$ & $ 58.6$ \nl
$-0.037$ & $-162.6$ & $ 64.0$ & $358.9$ & $123.8$ \nl
$-0.008$ & $-194.6$ & $ 60.4$ & $339.6$ & $ 97.8$ \tablebreak
$ 0.020$ & $-290.2$ & $ 79.1$ & $465.2$ & $116.6$ \nl
$ 0.048$ & $ -81.4$ & $ 75.1$ & $433.8$ & $135.3$ \nl
$ 0.076$ & $-125.0$ & $ 75.8$ & $368.1$ & $159.2$ \nl
$ 0.105$ & $-434.3$ & $110.6$ & $476.9$ & $152.5$ \nl
$ 0.133$ & $-307.3$ & $ 79.4$ & $382.3$ & $178.7$ \nl
$ 0.161$ & $-220.9$ & $ 62.6$ & $319.3$ & $110.6$ \nl
$ 0.190$ & $-252.6$ & $ 35.8$ & $174.8$ & $ 57.0$ \nl
$ 0.218$ & $-262.7$ & $ 25.6$ & $109.7$ & $ 49.7$ \nl
$ 0.246$ & $-238.7$ & $ 34.0$ & $205.1$ & $ 47.9$ \nl
$ 0.275$ & $-263.2$ & $ 38.6$ & $243.6$ & $ 68.7$ \nl
$ 0.303$ & $-275.7$ & $ 32.7$ & $195.4$ & $ 54.0$ \nl
$ 0.331$ & $-244.9$ & $ 29.4$ & $182.7$ & $ 51.4$ \nl
$ 0.360$ & $-212.9$ & $ 26.1$ & $173.4$ & $ 42.6$ \nl
\enddata
\end{deluxetable}

\clearpage

\begin{deluxetable}{ccccc}
\tablewidth{0pc}
\tablecaption{M31 Nuclear Kinematics --- Zoomed Pixel Resolution
\label{t.doublecols}}
\tablehead{\colhead{$R(\arcsec)$} & \colhead{$V$} &
\colhead{$\pm$} & \colhead{$\sigma$} &
\colhead{$\pm$}}
\startdata
$-0.929$ & $  82.1$ & $23.2$ & $169.3$ & $ 38.0$ \nl
$-0.872$ & $ 117.3$ & $23.0$ & $161.3$ & $ 37.3$ \nl
$-0.816$ & $ 151.0$ & $21.3$ & $145.9$ & $ 37.8$ \nl
$-0.759$ & $ 111.9$ & $23.0$ & $139.9$ & $ 38.4$ \nl
$-0.702$ & $ 140.8$ & $17.0$ & $105.8$ & $ 37.3$ \nl
$-0.646$ & $ 159.0$ & $14.5$ & $100.4$ & $ 31.3$ \nl
$-0.589$ & $ 191.8$ & $16.3$ & $147.4$ & $ 28.9$ \nl
$-0.532$ & $ 229.8$ & $22.5$ & $197.7$ & $ 39.7$ \nl
$-0.476$ & $ 202.4$ & $22.3$ & $186.7$ & $ 38.9$ \nl
$-0.419$ & $ 123.2$ & $23.7$ & $177.9$ & $ 39.6$ \nl
$-0.363$ & $ 188.9$ & $19.1$ & $124.7$ & $ 36.3$ \nl
$-0.306$ & $ 116.0$ & $26.3$ & $178.6$ & $ 49.3$ \nl
$-0.249$ & $  91.9$ & $21.0$ & $105.5$ & $ 42.7$ \nl
$-0.193$ & $  57.3$ & $31.2$ & $179.0$ & $ 48.5$ \nl
$-0.136$ & $ -43.2$ & $23.4$ & $177.9$ & $ 40.6$ \nl
$-0.079$ & $ -95.0$ & $32.1$ & $230.4$ & $ 44.4$ \nl
$-0.023$ & $-177.5$ & $49.3$ & $354.1$ & $ 78.8$ \nl
$ 0.034$ & $-159.6$ & $69.6$ & $462.5$ & $125.8$ \nl
$ 0.091$ & $-190.9$ & $70.4$ & $370.6$ & $146.0$ \nl
$ 0.147$ & $-247.9$ & $55.1$ & $340.6$ & $102.3$ \nl
$ 0.204$ & $-256.4$ & $27.0$ & $139.6$ & $ 45.6$ \nl
$ 0.261$ & $-245.2$ & $32.2$ & $218.5$ & $ 50.6$ \nl
$ 0.317$ & $-259.4$ & $27.1$ & $189.4$ & $ 45.7$ \nl
$ 0.374$ & $-223.5$ & $28.5$ & $213.2$ & $ 41.5$ \nl
\enddata
\end{deluxetable}

\clearpage

\end{document}